\documentclass[conference]{IEEEtran}
\IEEEoverridecommandlockouts

\usepackage{amssymb}
\usepackage{wrapfig}
\usepackage{rotating}
\usepackage{multirow}
\usepackage{array}
\usepackage{amssymb}
\usepackage{amsfonts}
\usepackage{graphicx}
\usepackage{epstopdf}
\usepackage{amsmath}
\usepackage{booktabs}
\usepackage{epigraph}
\usepackage{listings}
\usepackage{flushend}
\usepackage{url}
\usepackage{todonotes}
\usepackage{comment}
\usepackage[utf8]{inputenc}

\usepackage[noadjust]{cite}

\usepackage[numbers]{natbib}

\usepackage[bookmarksopen,bookmarksdepth=2]{hyperref}

\usepackage{color, colortbl}
\definecolor{Gray}{gray}{0.9}
\definecolor{White}{rgb}{255,255,255}

\newcommand{\MyPara}[1]{\vspace{.2em}\noindent\textit{\textbf{#1}}\hspace{.3em}}

\newcommand{\MyBox}[1]{\vspace{3mm}\noindent\framebox[\columnwidth][c]{\parbox[b]{0.95\columnwidth}{ #1 }}\vspace{3mm}}

\newboolean{showcomments}
\setboolean{showcomments}{true} 
\ifthenelse{\boolean{showcomments}}
 {
 
 }
 {\newcommand{\nbb}[2]{}
 
 }

\begin{document}

\title{Can I Solve It? Identifying APIs Required to Complete OSS Tasks
{}
}

\author{\IEEEauthorblockN{
Fabio Santos,\textsuperscript{$1$} 
Igor Wiese,\textsuperscript{$2$} 
Bianca Trinkenreich,\textsuperscript{$1$} 
Igor Steinmacher,\textsuperscript{$1,2$} 
Anita Sarma,\textsuperscript{$3$} 
Marco A. Gerosa\textsuperscript{$1$}}

\IEEEauthorblockA{\textsuperscript{$1$}\textit{Northern Arizona University, USA}, \textsuperscript{$2$}\textit{Universidade Tecnológica Federal do Paraná, Brazil},\textsuperscript{$3$} \textit{Oregon State University, USA}
\\ 
\\
fabio\_santos@nau.edu, igor@utfpr.edu.br, bianca\_trinkenreich@nau.edu, igorfs@utfpr.edu.br\\  anita.sarma@oregonstate.edu, marco.gerosa@nau.edu}
}

\maketitle
\begin{abstract}
Open Source Software projects add labels to open issues to help contributors choose tasks. However, manually labeling issues is time-consuming and error-prone.  Current automatic approaches for creating labels are mostly limited to classifying issues as a bug/non-bug. In this paper, we investigate the feasibility and relevance of labeling issues with the domain of the APIs required to complete the tasks. We leverage the issues' description and the project history to build prediction models, which resulted in precision up to 82\% and recall up to 97.8\%. We also ran a user study (n=74) to assess these labels' relevancy to potential contributors. The results show that the labels were useful to participants in choosing tasks, and the API-domain labels were selected more often than the existing architecture-based labels. Our results can inspire the creation of tools to automatically label issues, helping developers to find tasks that better match their skills.
\end{abstract}

\begin{IEEEkeywords}
API identification, Labelling, Tagging, Skills, Multi-Label Classification, Mining Software Repositories, Case Study
\end{IEEEkeywords}

\section{Introduction}



Finding tasks to contribute to in Open Source projects is challenging  ~\cite{wang2011bug,steinmacher2015understanding,steinmacher2015systematic,10.1145/2675133.2675215,stanik2018simple}. Open tasks vary in complexity and required skills, which can be difficult to determine solely by reading the task descriptions alone, especially for new contributors~\cite{zimmermann2010makes,Bettenburg:2007:QBR:1328279.1328284,vaz2019empirical}. Adding labels to the issues (a.k.a tasks, bug reports) help new contributors when they are choosing their tasks~\cite{steinmacher2018let}. However, community managers find that labeling issues is challenging and time-consuming \cite{9057411} because projects require skills in different languages, frameworks, databases, and Application Programming Interfaces (APIs). 

APIs usually encapsulate modules that have specific purposes (e.g., cryptography, database access, logging, etc.), abstracting the underlying implementation. If the contributors know which types of APIs will be required to work on each issue, they could choose tasks that better match their skills or involve skills they want to learn. 

Given this context, in this paper, we investigate the feasibility of automatically labeling issues with domains of APIs to facilitate contributors' task selection. Since an issue may require knowledge in multiple APIs, we applied a multi-label classification approach, which has been used in software engineering for purposes such as, classifying questions in Stack Overflow (e.g., \citet{xia2013tag}) and detecting types of failures (e.g., \citet{feng2018empirical}) and code smells (e.g., \citet{guggulothu2020code}). 

By employing an exploratory case study and a user study, we aimed to answer the following research questions:

\textbf{RQ1: To what extent can we predict the domain of APIs used in the code that fixes a software issue?} To answer RQ1, we employed a multi-label classification approach to predict the API-domain labels. We also explored the influence of task elements (i.e., title, body, and comments) and machine learning setup (i.e., n-grams and different algorithms) on the prediction model. Overall, we found that pre-processing the issue body using unigram and Random Forest algorithm can predict the API-domain labels with up to 82\% precision and up to 97.8\% of recall. This configuration outperformed recent approaches reported in the literature~\cite{el2020automatic}.  

\textbf{RQ2. How relevant are the API-domain labels to new contributors?} To answer RQ2, we conducted a study with 74 participants from both academia and industry. After asking participants to select and rank real issues they would like to contribute to, we provided a follow-up survey to determine what information was relevant to make the decision. We compared answers from the treatment group (with access to the API-domain labels) with the control group (who used only the pre-existing project labels). The participants considered API-domain labels more relevant than the project labels---with a large effect size.

These results indicate that labeling issues with API domain is feasible and relevant for new contributors who need to determine which issues to contribute.

\section{Related Work}
\label{sec:related}

New contributors need specific guidance on what to contribute~\cite{Park.Jensen_2009,steinmacher2018let}. In particular, finding an appropriate issue can be a daunting task, which can discourage contributors~\cite{steinmacher2015understanding}. Social coding platforms like GitHub\footnote{\href{http://bit.ly/NewToOSS}{http://bit.ly/NewToOSS}} encourages projects to label issues that are easy for new contributors, which is done by several communities (e.g.  LibreOffice,\footnote{\url{https://wiki.documentfoundation.org/Development/EasyHacks}} KDE,\footnote{\url{https://community.kde.org/KDE/Junior_Jobs}} and Mozilla\footnote{\url{https://wiki.mozilla.org/Good_first_bug}}) However, community managers argue that manually labeling issues is difficult and time-consuming~\cite{9057411}.

Several studies have proposed ways to automatically label issues as bug/non-bug, combining text mining techniques with classification to mitigate this problem. For example, \citet{antoniol2008bug} compared text-based mining with Naive Bayes (NB), Logistic Regression (LR), and Decision Trees (DT) to process data from titles, descriptions, and discussions and achieved a recall up to 82\%. \citet{pingclasai2013classifying} used the same techniques to compare a topic and word-based approach and found F-measures from 0.68 to 0.81 using the topic-based approach. More recently, \citet{zhou2016combining} used two-stage processing, introducing the use of structured information from the issue tracker, improving the recall obtained by \citet{antoniol2008bug}. \citet{kallis2019ticket} simplified the data mining step to produce a tool able to classify issues on demand. They used the title and body to create a bag of words used to classify issues as ``bug report'', ``enhancement'', and ``question''. \citet{el2020automatic} applied type detection on issues and attempted to transfer learning to other projects using the same training data. The best results had F-Measures around 0.64 - 0.75. Finally, \citet{xia2013tag} employed a multi-label classification using text data from Stack Overflow questions, obtaining recall from 0.59 to 0.77. 

As opposed to these related work---which focuses mostly on classifying the type of issue (e.g., bug/non-bug)---our work focuses on identifying the domain of APIs used in the implementation code, which might reflect skills needed to complete a task. 

Regarding APIs, recent work focuses on understanding the crowd's opinion about API usage \cite{8643972}, understanding and categorizing the API discussion \cite{6676877}, creating tutorial sections that explain a given API type \cite{7194633}, generating/understanding API documentation \cite{7886920,7985647}, providing API recommendations \cite{10.1145/3238147.3238191,8186224, 8478004}, offering API tips to developers \cite{8816774}, and defining the skill space for APIs, developers, and projects \cite{dey2020representation}. In contrast to these previous work, we focus on predicting the domain of the API used in the code that fixes an issue.

\section{Method}

\begin{figure}[!htb]
\centering
\includegraphics[width=.48\textwidth] {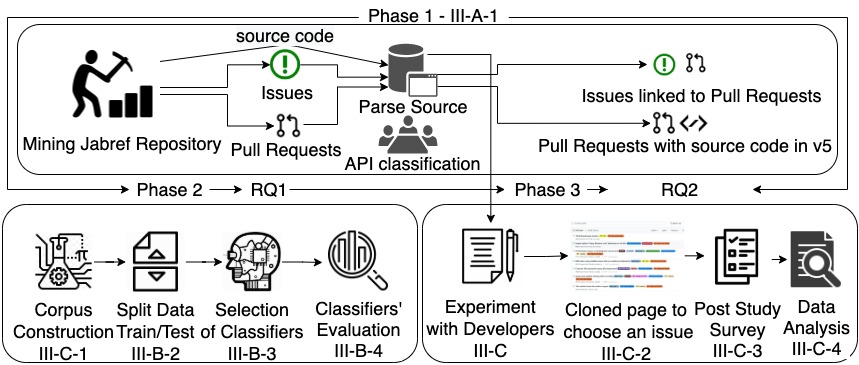}
\caption{Research Design}
\label{fig:researchdesign}
\end{figure}

This study comprises three phases, as summarized in Fig.~\ref{fig:researchdesign}: mining software repository, building the classifiers, and evaluating the API-domain labels with developers. To foster reproducibility, we provide a publicly available dataset\footnote{\url{http://doi.org/10.5281/zenodo.4628599}} containing the raw data, the Jupyter notebook scripts that build and test the models, and the survey data. 

We conducted an exploratory case study \cite{CaseStudyBook} using the JabRef project~\cite{JabRef} as our case study. JabRef is an open-source bibliography reference manager developed by a community of volunteers, including contributors with different background and diverse set of skills (with and without computer science background)---this helped us evaluate the approach with a diverse set of contributor profiles. JabRef is a mature and active project created in 2003 (migrated to GitHub in 2014), with 15.7k commits, 42 releases, 337 contributors, 2.7k closed issues, and 4.1k closed pull requests. JabRef has also been frequently investigated in scientific studies~\cite{olsson2017relationship,mayr2014benchmarking,herold2020initial,shi2014empirical,feyer2017integration}. We chose JabRef as our case study because of these characteristics and because we have access to the project's contributors.

\subsection{Phase 1 - Mining JabRef Repository}
\label{sec:DataCollection}

We used the GitHub API to collect data from JabRef. We collected 1976 issues and pull requests (PR), including title, description (body), comments, and submission date. We also collected the name of the files changed in the PR and the commit message associated with each commit. The data was collected in April 2020. 

After collecting the data, we filtered out open issues and pull requests not explicitly linked to issues. To find the links between pull requests and issues, we searched for the symbol \verb|#|issue\verb|_|number in the pull request title and body and checked the URL associated with each link. We manually inspected a random sample of issues to check whether the data was correctly collected and reflected what was shown on the GitHub interface. Two authors manually examined 50 issues, comparing the collected data with the GitHub interface. All records were consistent. We also filtered out issues linked to pull requests without at least one Java file (e.g., those associated only with documentation files). Our final dataset comprises 705 issues and their corresponding pull requests.



We also wrote a parser to process all Java files from the project. In total, 1,692 import statements from 1,472 java sources were mapped to 796 distinct APIs. The parser identified all classes, including the complete namespace from each import statement. Then we filtered out APIs not found in the latest version of the source code (JabRef 5.0) to avoid recommending APIs that were no longer used in the project. 

Then, we employed a card-sorting approach to manually classify the imported APIs into higher-level categories based on the API's domain. For instance, we classified ``java.nio.x'' as ``IO', ``java.sql.x'' as ``Database'', and ``java.util.x'' as ``Utils'''. A three-member team performed this classification (first, second, and forth authors), one of them is a contributor of JabRef and another one is an expert Java developer. They analyzed the APIs identified in the previous step and each person individually classified the API and discussed to reach a consensus. After classifying all the APIs, the researchers conducted a second round to revise the classification ($\sim$8 hours). During this step, we excluded some labels and aggregated some others. The final set of categories of API domains contains: Google Commons, Test, OS, IO, UI, Network, Security, OpenOffice Documents, Database, Utils, PDF, Logging, and Latex. We used these categories as labels to the 705 issues previously collected based on the presence of the corresponding APIs in the changed files. We used this annotated set to build our training and test sets for the multi-label classification models.

\subsection{Phase 2 - Building the Multi-label Classifiers}
\label{sec:MultilabelClassification}

\subsubsection{Corpus construction}
\label{sec:DataExtraction}

To build our classification models, we created a corpus comprising issue titles, body, and comments. We converted each word to lowercase and removed URLs, source code, numbers, and punctuation. After that, we removed stop-words and stemmed the words using the Python nltk package. We also filtered issue and pull request templates\footnote{\href{http://bit.ly/NewToOSS}{http://bit.ly/NewToOSS}} since the templates were not consistently used among all the issues. We found in our exploratory studies that their repetitive structure introduced too much noise.

Next, similar to other studies~\cite{ramos2003using,behl2014bug,vadlamani2020studying}, we applied TF-IDF, which is a technique for quantifying word importance in documents by assigning a weight to each word. After applying TF-IDF, we obtained a vector for each issue. The vector length is the number of terms used to calculate the TF-IDF plus the 13 labels in the dataset. Each label received a binary value (0 or 1), indicating whether the corresponding API-domain is present in the issue and each term received the TF-IDF score.


\subsubsection{Training/Test Sets}
\label{sec:TrainTest}

We split the data into training and test sets using the ShuffleSplit method~\cite{MultilabelBook}, which is a model selection technique that emulates cross-validation for multi-label classifiers. We randomly split our 705 issues into a training set with 80\% (564) of the issues and a test set with the remaining 20\% (142 issues). To avoid overfitting, we ran each experiment ten times, using ten different training and test sets to match a 10-fold cross validation. To improve the balance of the data set, we ran the SMOTE algorithm for multi-label approach~\cite{charte2015mlsmote}.

\subsubsection{Classifiers}
\label{sec:Classsifier}

To create the classification models, we chose five classifiers that work with the multi-label approach and implement different strategies to create learning models: Decision Tree, Random Forest (ensemble classifier), MPLC Classifier (neural network multilayer perceptron), MLkNN (multi-label lazy learning approach based on the traditional K-nearest neighbor algorithm)~\cite{zhang2007ml,MultilabelBook}, and Logistic Regression. We ran the classifiers using the Python sklearn package and tested several parameters. For the RandomForestClassifier, the best classifier (see Section~\ref{sec:results}), we kept the following parameters: $criterion='entropy'$ ,$max\_depth=50$, $min\_samples\_leaf=1$, $min\_samples\_split=3$,$n\_estimators=50$.

\subsubsection{Classifiers Evaluation}
\label{sec:Evaluation}

To evaluate the classifiers, we employed the following metrics (also calculated using the scikit-learn package):

\begin{itemize}

\item \textbf{Hamming loss} measures the fraction of the wrong labels to the total number of labels.

\item \textbf{Precision} measures the proportion between the number of correctly predicted labels and the total number of predicted labels. 
 
\item \textbf{Recall} corresponds to the percentage of correctly predicted labels among all truly relevant labels. 
 
\item \textbf{F-measure} calculates the harmonic mean of precision and recall. F-measure is a weighted measure of how many relevant labels are predicted and how many of the predicted labels are relevant.
 
\item \textbf{Matthews correlation coefficient - MCC} calculates the Pearson product-moment correlation coefficient between actual and predicted values. It is an alternative measure unaffected by the unbalanced dataset issue~\cite{chicco2020advantages}.
\end{itemize}

\subsubsection{Data Analysis}
\label{sec:DataAnalysis}




To conduct the data analysis, we used the aforementioned evaluation metrics and the confusion matrix logged after each model's execution. We used the Mann-Whitney U test to compare the classifier metrics, followed by Cliff's delta effect size test. The Cliff's delta magnitude was assessed using the thresholds provided by \citet{Romano:2006}, i.e. $|$d$|$$<$0.147 ``negligible'', $|$d$|$$<$0.33 ``small'', $|$d$|$$<$0.474 ``medium'', otherwise ``large''. 


\subsubsection{Dataset Analysis}
\label{sec:DatasetDescription}

Multi-label datasets are usually described by label cardinality and label density~\cite{MultilabelBook}. Label cardinality is the average number of labels per sample. Label density is the number of labels per sample divided by the total number of labels, averaged over the samples. For our dataset, the label cardinality is 3.04. The density is 0.25. These values consider the 705 distinct issues and API-domain labels obtained after the previous section's pre-processing steps. Since our density can be considered high, the multi-label learning process or inference ability is not compromised~\cite{blanco2019multi}.

For the remainder of our analysis, we removed the API label ``Utils,'' since we found that this label was present in 96\% of the issues in our final dataset and has an overly generic meaning. The API-domain labels ``IO'', ``UI'', and ``Logging" had 492, 452, and 417 occurrences respectively. These last three labels occurred in approximately 60\% of the issues. We also observed that ``Test'', ``Network'', and ``Google Commons'' appeared in almost 29\% of the issues (212, 208, and 206 times). ``SO'', ``Database'', ``PDF'', ``Open Office'', ``Security'', and ``Latex'' were less common, with 56, 31, 21, 21, 20, and 14 occurrences respectively. 

\begin{figure}[!hbt]
\centering
\includegraphics[width=.45\textwidth, trim= 50px 45px 50px 40px] {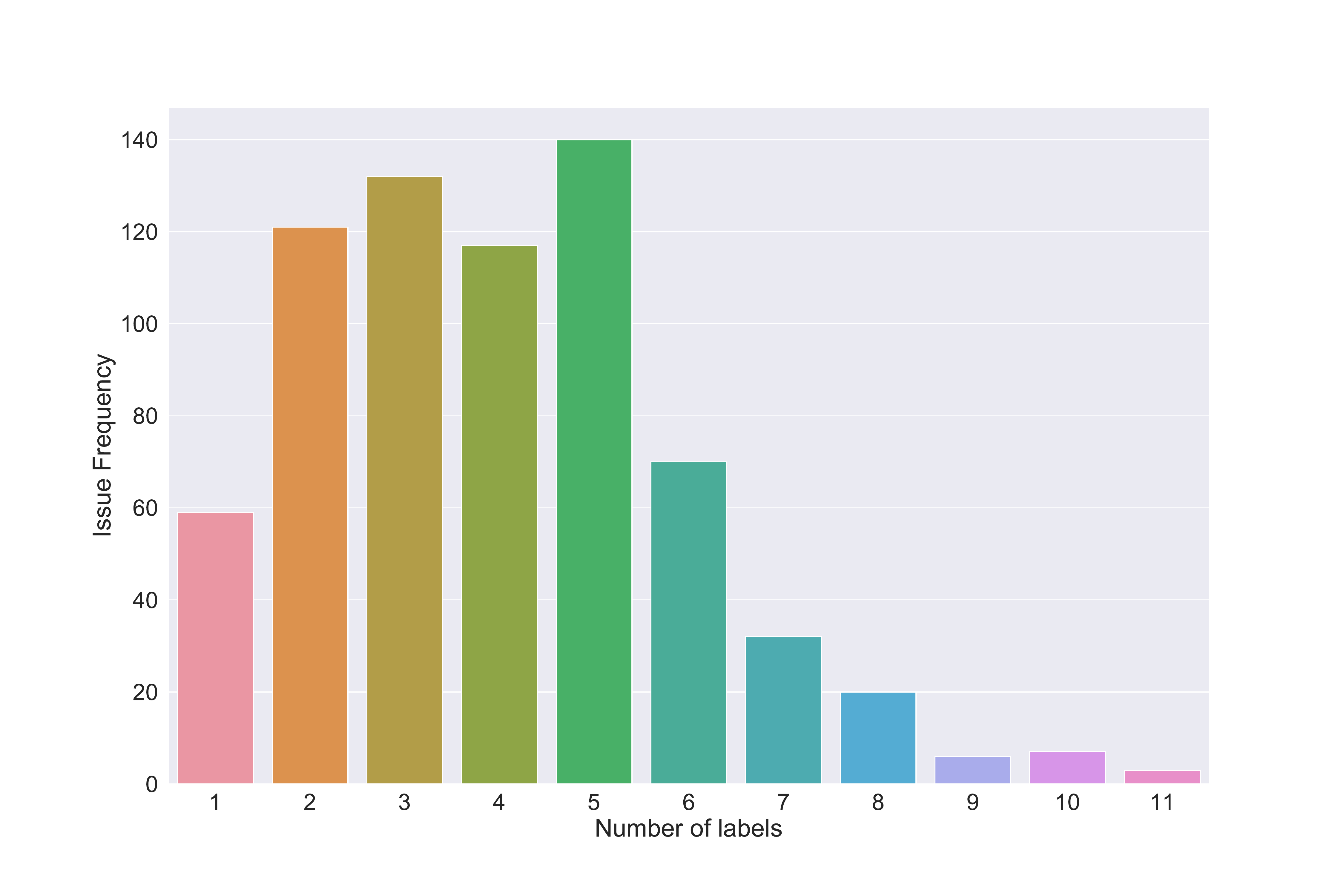}
\caption{Number of labels per issue}
\label{fig:issueFrequency}
\end{figure}

Finally, we checked the distribution of the number of labels per issue (Fig. \ref{fig:issueFrequency}). We found 140 issues with five labels, 132 issues with three labels, 121 issues with two labels, and 117 issues with four labels. Only 8.5\% of issues have one label, which confirms a multi-label classification problem.

\subsection{Phase 3 - Evaluating the API-Domain Labels with Developers}
\label{sec:QualitativeExp}

To evaluate the relevancy of the API-domain labels from a new contributor's perspective, we conducted an experimental study with 74 participants. We created two versions of the JabRef issues page (with and without our labels) and divided our participants into two groups (between-subjects design). We asked participants to choose and rank three issues they would like to contribute and answer a follow-up survey about what information supported their decision. The artifacts used in this phase are also part of the replication package.

\subsubsection{Participants}
\label{sec:Participants}


We recruited participants from both industry and academia. We reached out to our own students in addition to instructors and IT managers of our personal and professional networks and asked them to help in inviting participants. From industry, we recruit participants from one medium-size IT startup hosted in Brazil and the IT department of a large and global company. Students included undergraduate and graduate computer science students from one university in the US and two others in Brazil. We also recruited graduate data science students from a university in Brazil, since they are also potential contributors to the JabRef project. We present the demographics of the participants in Table \ref{tab:demographics}. We offered an Amazon Gift card to incentivize participation.

We categorized the participants' development tenure into novice and experienced coders, splitting our sample in half---below and above the average ``years as professional developer'' (4). We also segmented the participants between industry practitioners and students. Participants are identified by: "P", followed by a sequential number and a character representing the location where they were recruited (University: U \& Industry: I); "T" for Treatment and "C" for Control groups.

\begin{table}[!htb]
\caption{Demographics Subgroups for the Experiment's Participants}

\centering
\label{tab:demographics}
\begin{tabular}{lrrllrr}
\cline{1-3} \cline{5-7}
\textbf{Popu-}     & \textbf{Quan-} & \textbf{Percent-} &  & \textbf{Tenure} & \textbf{Quan-} & \textbf{Percent-} \\ 
\textbf{lation}     & \textbf{tity} & \textbf{age} &  & \textbf{} & \textbf{tity} & \textbf{age} \\ 
\cline{1-3} \cline{5-7} 
Industry & 41                     & 55.5                   &  & Expert & 19                     & 25.7                     \\

Student  & 33                     & 44.5                   &  & Novice       & 55                     & 74.3                     \\ 

\cline{1-3} \cline{5-7} 
\end{tabular}
\end{table}

The participants were randomly split into two groups: Control and Treatment. From the 120 participants that started the survey, 74 (61.7\%) finished all the steps, and we only considered these participants in the analysis. We ended up with 33 and 41 participants in the Control and Treatment groups, respectively. 

\subsubsection{Experiment Planning}
\label{sec:ExperimentPlanning}

We selected 22 existing JabRef issues and built mock GitHub pages for Control and Treatment groups. The issues were selected from the most recent ones, trying to maintain similar distributions of the number of API-domain labels predicted per issue and the counts of predicted API-domain labels (see Section \ref{sec:DatasetDescription}). The control group mocked page had only the original labels from the JabRef issues and the treatment group mocked page presented the original labels in addition to those API-domain labels. These pages are available in the replication package.

\subsubsection{Survey Data Collection}
\label{sec:SurveyDataCollection}

\begin{figure}
\centering
\includegraphics[width=.45\textwidth]{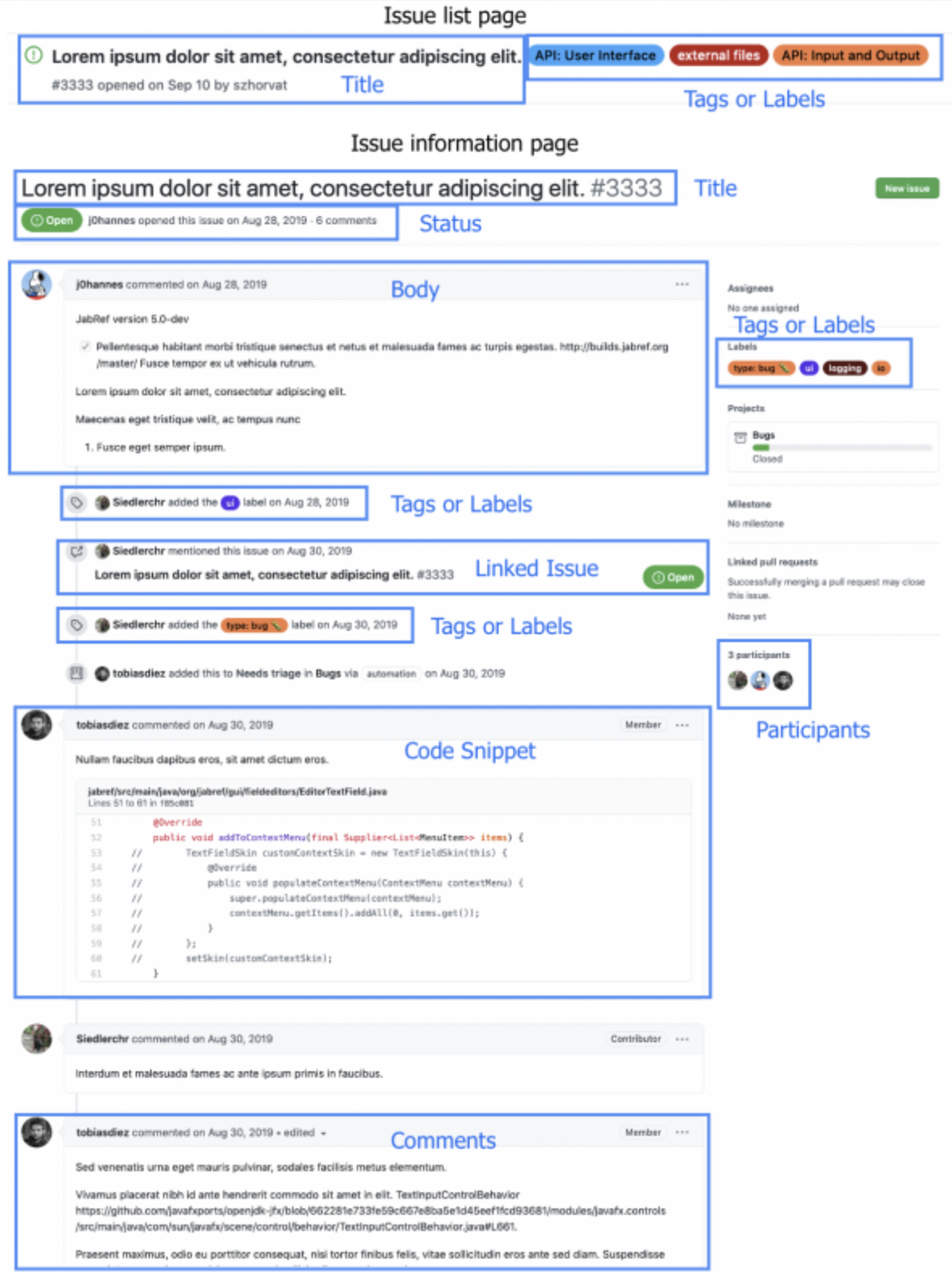}
\caption{Survey question about the regions relevance}
\label{fig:hotspotsurvey}
\end{figure}

The survey included the following questions/instructions:

\begin{itemize}
    \item Select the three issues that you would like to work on.
    \item Select the information (region) from the issue page that helped you deciding which issues to select (Fig: \ref{fig:hotspotsurvey}).
    \item Why is the information you selected relevant? (open-ended question)
    \item Select the labels you considered relevant for choosing the three issues
\end{itemize} 

The survey also asked about participants' experience level, experience as an OSS contributor, and expertise level in the technologies used in JabRef.


Fig. \ref{fig:hotspotsurvey} shows an example of an issue details page and an issue entry on an issue list page. After selecting the issues to contribute, the participant was presented with this page to select what information (region) was relevant to the previous issue selection.

\subsubsection{Survey Data Analysis}
\label{sec:SurveyDataAnalysis}


Next, to understand participants' perceptions about what information (regions) they considered important and the relevancy of the API-domain labels, we first compared treatment and control groups' results. We used violin plots to visually compare the distributions and measured the effect size using the Cliff's Delta test.

Then, we analyzed the data, aggregating participants according to their demographic information, resulting in the subgroups presented in Table~\ref{tab:demographics}. We calculated the odds ratio to check how likely it would be to get similar responses from both groups. We used a 2x2 contingency table for each comparison---for instance, industry practitioners vs. students and experienced vs. novice coders. We used the following formula to calculate the odds ratio~\cite {szumilas2010explaining}: 
\begin{center}
\vspace{2mm}
$Odds Ratio (OR) = \frac{(a/c)}{(b/d)}$
\vspace{2mm}
\end{center}
Probabilities $>$ 1 mean that the first subgroup is more likely to report a type of label, while probabilities less than 1 mean that the second group has greater chances (OR)~\cite {odds-ratio}.

To understand the rationale behind the label choices, we qualitatively analyzed the answers to the open question ("Why was the information you selected relevant?"). We selected representative quotes to illustrate the participants' perceptions about the labels' relevancy. 




\section{Results}\label{sec:results}

We report the results grouped by research question.

\subsection{RQ1. To what extent can we predict the domain of APIs used in the code that fixes a software issue?}


To predict the API domains identified in the files changed in each issue (RQ1), we started by testing a simple configuration used as a baseline. For this baseline model, we used only the issue \textsc{title} as input and the Random Forest (RF) algorithm, since is insensitive to parameter settings~\cite{RandomForestShane} and is usually yields good results in software engineering studies \cite{petkovic2016using,goel2017random,pushphavathi2014novel,satapathy2016early}. Then, we evaluated the corpus configuration alternatives, varying the input information: only \textsc{title} (T), only \textsc{body} (B), \textsc{title} and \textsc{body}, and \textsc{title}, \textsc{body}, and \textsc{comments}. To compare the different models, we selected the best Random Forest configuration and used the Mann-Whitney U test with the Cliff's-delta effect size.

We also tested alternatives configurations using n-grams. For each step, the best configuration was kept. Then, we used different machine learning algorithms comparing with a dummy (random) classifier.


\begin{table}[ht!]
 \begin{center}
 \caption{overall metrics (section III-B-4) from models created to evaluate the corpus. Hla - Hamming Loss }
 \label{tab:results}
 \begin{tabular}{c|r|r|r|r} 
  \hline

  \textbf{Model} & \textbf{Precision} & \textbf{Recall} & \textbf{F-measure} &
  \textbf{Hla} \\
  \hline
Title (T) 
&0.717 &	0.701 &	0.709 & 0.161\\
Body (B)
& 0.752 &	0.742 &	0.747 & 0.143\\
T, B 
& 0.751 &	0.738 &	0.744 & 0.145\\
T, B, Comments 
& 0.755	& 0.747 &	0.751 & 0.142\\
 \hline

 \end{tabular}
 \end{center}
\end{table}

As can be seen in Table~\ref{tab:results}, when we tested different inputs and compared to \textsc{Title} only, all alternative settings provided better results. We could observe improvements in terms of precision, recall, and F-measure. When using \textsc{title}, \textsc{body}, and \textsc{comments}, we reached Precision of 75.5\%, Recall of 74.7\%, and F-Measure of 75.1\%. 



\begin{table}[h!]
 \begin{center}
 \caption{Cliff's Delta for F-Measure and Precision: comparison of corpus models alternatives - Section III-B-1. Title(T), Body(B) and Comments (C).}
 \label{tab:resultsH1H2H3cliffs}
 \begin{tabular}{l|r|l|r|l} 
 \hline
 \textbf{Corpus} & \multicolumn{4}{c}{\textbf{Cliff's delta}} \\
 \cline{2-5}
 \textbf{Comparison} & \multicolumn{2}{c|}{\textbf{F-measure}} & \multicolumn{2}{c}{\textbf{Precision}}\\
  \hline

T versus B & -0.86 & large***  & -0.92 & large***\\
T versus T+B & -0.8 & large** & -0.88 & large***\\
T versus T+B+C & -0.88 & large** & -0.88 & large*** \\
B versus T+B & 0.04 & negligible & 0.04 & negligible \\
B versus T+B+C & -0.24 & small & -0.12 & negligible \\
T+B versus T+B+C & -0.3 & small & -0.08 & negligible\\
\hline
\multicolumn{5}{l}{\scriptsize\textit{* p $\leq$ 0.05;
** p $\leq$ 0.01; 
*** p $\leq$ 0.001}}
\\

 \end{tabular}
 \end{center}
\end{table}

We found statistical differences comparing the results using \textsc{title} only and all the three other corpus configurations for both F-measure (p-value $\leq$ 0.01 for all cases, Mann-Whitney U test) and precision (p-value $\leq$ 0.001 for all cases, Mann-Whitney U test) with large effect size. \textsc{Title}+\textsc{body}+\textsc{comments} performed better than all others in terms of precision, recall, and F-measure. However, the results suggest that using only the \textsc{body} would provide good enough outcomes, since there was no statistically significant difference comparing to the other two configurations---using \textsc{title} and/or \textsc{comments} in addition to the \textsc{body}--- and it achieved similar results with less effort. The model built using only \textsc{body} presented only 14.3\% incorrect predictions (hamming loss metric) for all 12 labels. Table \ref{tab:resultsH1H2H3cliffs} shows the Cliff's-delta comparison between each pair of corpus configuration. 



\begin{figure}
\centering
\includegraphics[width=.5\textwidth, trim= 0px 25px 20px 40px]{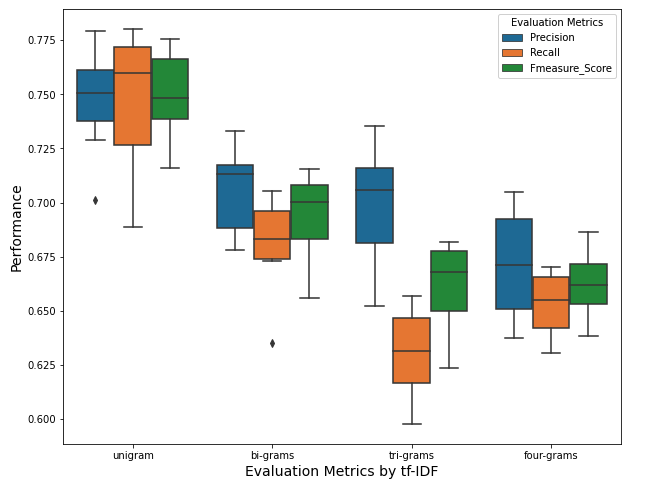}
\caption{Comparison between the unigram model and n-grams models}
\label{fig:baselineH5iftdf}
\end{figure}


Next, we investigated the use of bi-grams, tri-grams, and four-grams comparing the results to the use of unigrams. We used the corpus with only issue \textsc{body} for this analysis, since this configuration performed well in the last step. Fig. \ref{fig:baselineH5iftdf} and Table \ref{tab:resultsH5cliffs} present how the Random Forest model performs for each n-gram configuration. The unigram configuration outperformed the others with large effect size. 




\begin{table}[h!]
 \begin{center}
 \caption{Cliff's Delta for F-Measure and Precision: Comparison between n-grams models - Section III-B-5}
 \label{tab:resultsH5cliffs}
 \begin{tabular}{l|r|l|r|l} 
 \hline
 \textbf{n-Grams} & \multicolumn{4}{c}{\textbf{Cliff's delta}} \\
 \cline{2-5}
 \textbf{Comparison} & \multicolumn{2}{c|}{\textbf{F-measure}} & \multicolumn{2}{c}{\textbf{Precision}}\\
  \hline

1 versus 2 & 1.0 & large*** & 0.86 & large***\\
1 versus 3 & 1.0 & large*** & 0.84 & large*** \\
1 versus 4 & 1.0 & large*** & 0.96 & large*** \\
2 versus 3 & 0.8 & large** & 0.18 & small \\
2 versus 4 & 0.78 & large** & 0.72 & large** \\
3 versus 4 & 0.12 & negligible & 0.62 & large*\\
\hline
\multicolumn{5}{l}{\scriptsize\textit{* p $\leq$ 0.05;
** p $\leq$ 0.01; 
*** p $\leq$ 0.001}}
\\
 \end{tabular}
 \end{center}
\end{table}

\begin{figure}[!hbt]
\centering
\includegraphics[width=.5\textwidth, trim= 10px 25px 20px 20px]{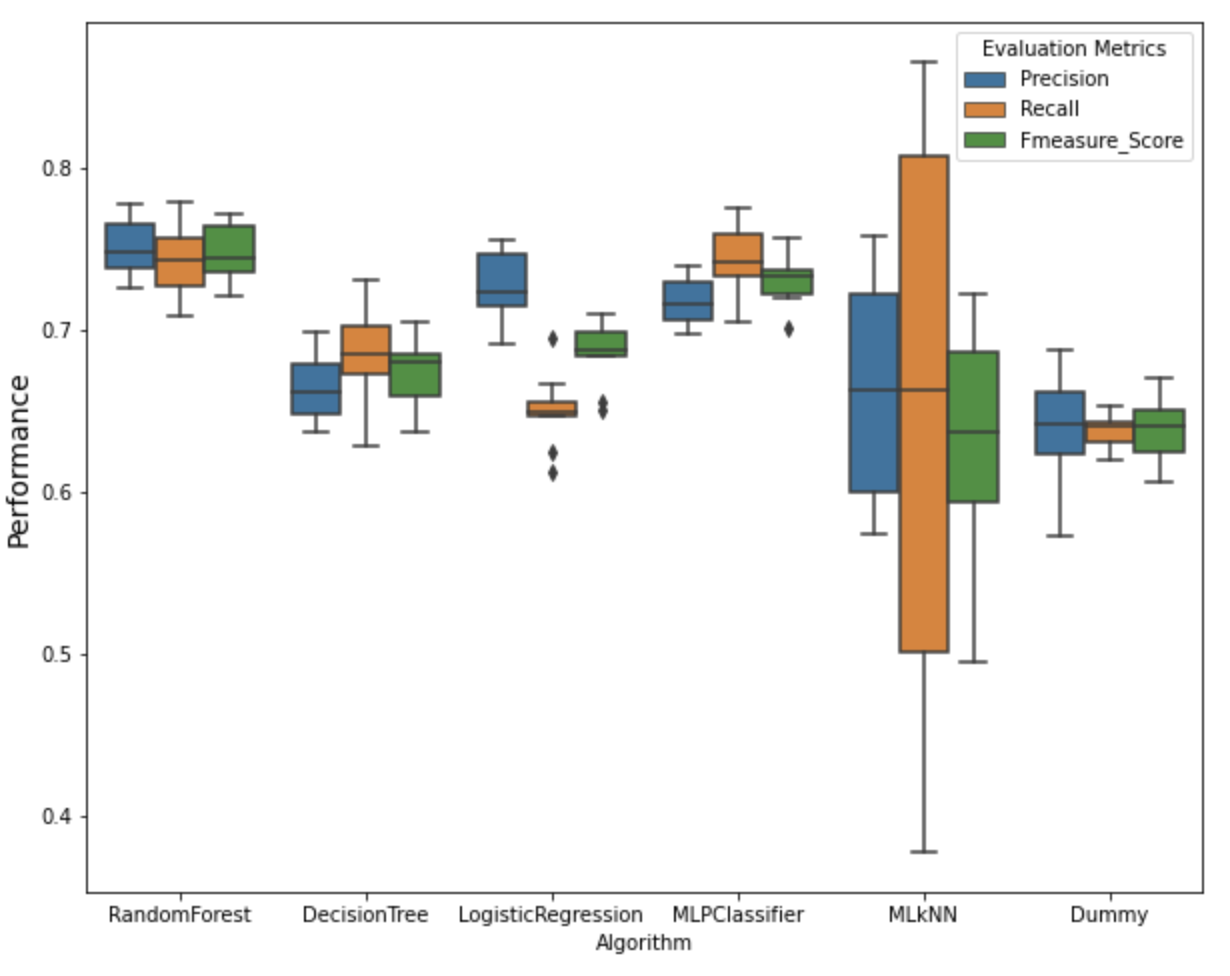}
\caption{Comparison between the baseline model and other machine learning algorithms}
\label{fig:baselineH6}
\end{figure}

Finally, to investigate the influence of the machine learning (ML) classifier, we compared several options using the title with unigrams as a corpus: Random Forest (RF), Neural Network  Multilayer Perceptron (MLPC), Decision Tree (DT), LR, MlKNN, and a Dummy Classifier with strategy ``most\_frequent''. Dummy or random classifiers are often used as baseline \cite{saito2015precision, flach2015precision}. We used the implementation from the Python package scikit-learn~\cite{sklearn}. Fig. \ref{fig:baselineH6} shows the comparison among the algorithms, and Table \ref{tab:resultsH6cliffs} presents the pair-wise statistical results comparing F-measure and precision using Cliff's delta. 

\begin{table}[hbt!]
 \begin{center}
 \caption{Cliff's Delta for F-Measure and Precision: Comparison between machine learning algorithms - Section III-B-5}
 \label{tab:resultsH6cliffs}
 \begin{tabular}{l|r|l|r|l} 
  \hline
\textbf{Algorithms} & \multicolumn{4}{c}{\textbf{Cliff's delta}} \\
 \cline{2-5}
\textbf{Comparison}  & \multicolumn{2}{c|}{\textbf{F-measure}} & \multicolumn{2}{c}{\textbf{Precision}}\\
  \hline

RF versus LR & 1.0 & large*** & 0.62  & large*\\
RF versus MLPC & 0.54  & large* & 0.88  & large*** \\
RF versus DT & 1.0  & large*** & 1.0  & large*** \\
RF versus MlkNN & 0.98  & large*** & 0.78  & large*** \\
LR versus MLPC & -0.96  & large*** & 0.24  & small \\
LR versus DT & 0.4  & medium & 0.94  & large*** \\
LR versus MlkNN & 0.5  & large* & 0.48  & large* \\
MPLC versus DT & 0.98  & large*** & 0.98  & large*** \\
MPLC vs. MlkNN & 0.94  & large*** & 0.32  & small \\
MlkNN versus DT & -0.28  & small & 0.0  & negligible \\
RF versus Dummy & 1.0 & large*** &  1.0 & large*** \\
\hline
\multicolumn{5}{l}{\scriptsize\textit{* p $\leq$ 0.05;
** p $\leq$ 0.01; 
*** p $\leq$ 0.001}}
\\
 \end{tabular}
 \end{center}
\end{table}


Random Forest (RF) and Neural Network Multilayer Perceptron (MLPC) were the best models when compared to Decision Tree (DT), Logistic Regression (LR), MlKNN, and Dummy algorithms. Random Forest outperformed these four algorithms with large effect sizes considering F-measure and precision.


\MyBox{\textbf{\emph{RQ1 Summary.}} It is possible to predict the API-domain labels with precision of 0.755, recall of 0.747, F-measure of 0.751, and 0.142 of Hamming loss using the Random Forest algorithm, \textsc{title}, \textsc{body} and \textsc{comments} as the corpus, and unigrams.} 



\subsection{RQ2. How relevant are the API-domain labels to new contributors?}

To answer this research question, we conducted an experiment with 74 participants and analyzed their responses. 

\begin{figure}[!hbt]
\centering
\includegraphics[width=.4 \textwidth] {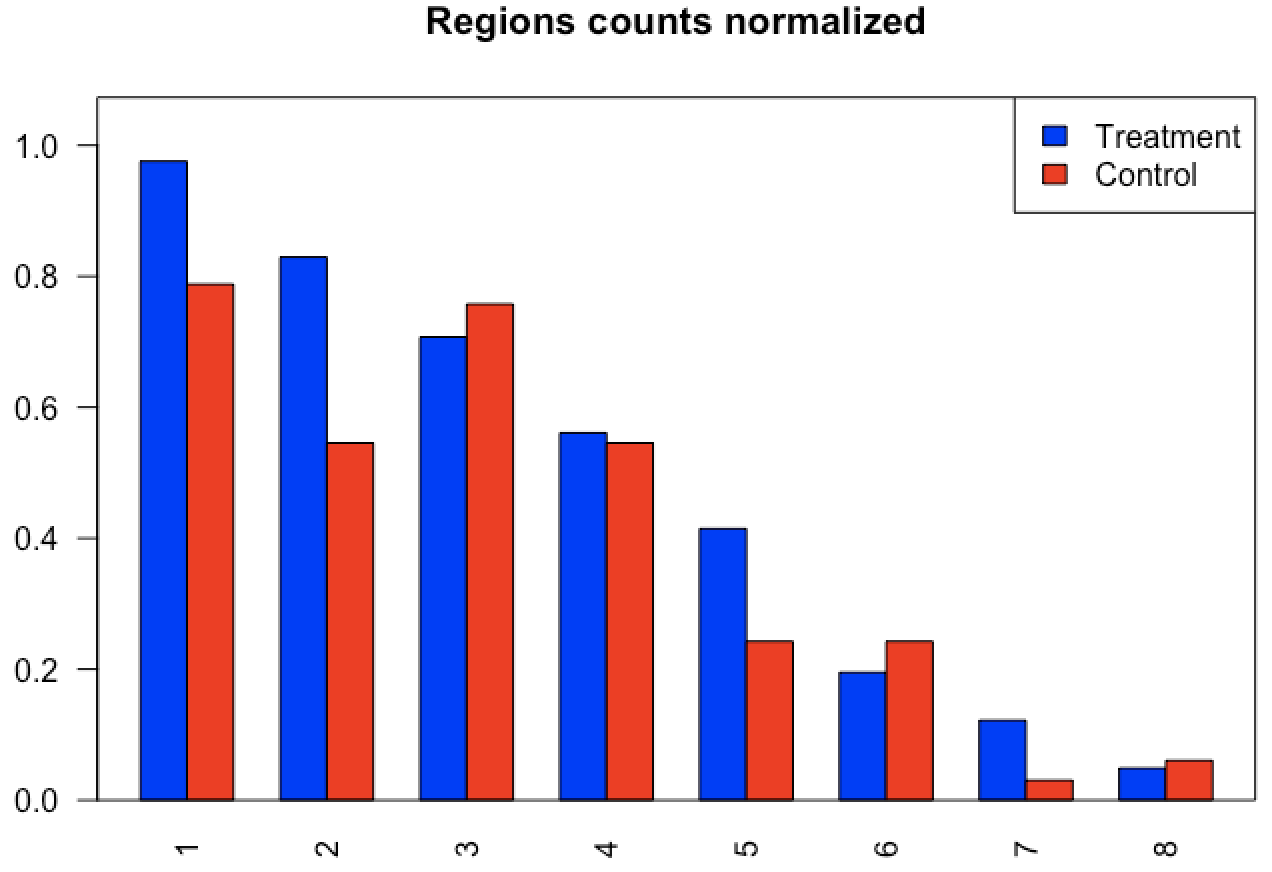}
\caption{The regions counts (normalized) of the issue's information page selected as most relevant by participants from Treatment and Control groups. 1-Title,2-Label,3-Body,4-Code,5-Comments,6-Author,7-Linked issues,8-Participants.}
\label{fig:hotmapchoicesTC}
\end{figure}

\textbf{What information is used when selecting a task?}
Understanding the type of information that participants use to make their decision while selecting an issue to work on can help projects better organize such information on their issue pages. Fig.~\ref{fig:hotmapchoicesTC} shows the different regions that participants found useful. In the control group, the top two regions of interest included the body of the issue (75.7\%) and the title (78.7\%), followed by the labels (54.5\%) and then the code itself (54.5\%). This suggests that the labels generated by the project were only marginally useful and participants had to also review the code. In contrast, in the Treatment group, the top four regions of interest by priority were: Title, Label, Body, and then Code (97.5\%, 82.9\%, 70.7\%, 56.1\%, respectively). This shows that participants in the Treatment group found the labels more useful than those participants in the Control group: 82.9\% usage in the Treatment group as compared to 54.5\% in the Control group. Comparing the body and the label regions in both groups, we found that participants from the Treatment group selected 1.6x more labels than the Control groups (p=0.05). The odds ratio analysis suggests that labels were more relevant in the Treatment groups.


Qualitative analysis of the reason behind the choice of participants in the Treatment group reveals that the Title and the Labels together provided a comprehensive view of the issue. For instance, P4IT mentioned: \textit{"labels were useful to know the problem area and after reading the title of the issues, it was the first thing taken into consideration, even before opening to check the details"}. Participants found the labels to be useful in pointing out the specific topic about the issue, as P14IT stated: \textit{``[labels are] hints about what areas have connection with the problem occurring''}.



\textbf{What is the role of labels?}
We also investigated which type of labels helped the participants in their decision making. We divide the labels available to our participants into three groups based on the type of information they imparted. 

\begin{itemize}
\item Issue type (already existing in the project): This included information about the type of the task: Bug, Enhancement, Feature, Good First Issue, and GSoC.
\item Code component (already existing in the project): This included information about the specific Code components of JabRef: Entry, Groups, External.Files, Main Table, Fetcher, Entry.Editor, Preferences, Import, Keywords
\item API-domain (new labels): the labels that were generated by our classifier (IO, UI, Network, Security, etc.). These labels were available only to the Treatment group. 
\end{itemize} 

\begin{table}[htb]
\centering

\caption{label distributions among the control and treatment groups}
\label{tab:distributionLabels}
\begin{tabular}{lllll}
\hline
Type of Label & Control                 & C \%                                          & Treatment               & T \%                                          \\ \hline
Issue Type          & \multicolumn{1}{r}{145} & \multicolumn{1}{r}{56.4}                      & \multicolumn{1}{r}{168} & \multicolumn{1}{r}{36.8}                      \\
Components    & \multicolumn{1}{r}{112} & \multicolumn{1}{r}{43.6} & \multicolumn{1}{r}{94}  & \multicolumn{1}{r}{20.6} \\
API Domain          & \multicolumn{1}{r}{-}   & \multicolumn{1}{r}{-}                                             & \multicolumn{1}{r}{195} & \multicolumn{1}{r}{42.7}                      \\ \hline
 
\end{tabular}
\end{table}

Table \ref{tab:distributionLabels} compares the labels that participants considered relevant (section III-C-3) across the Treatment and Control groups, distributed across these label types. In the Control group, a majority of selected labels (56.4\%) relate to the type of issue (e.g., Bug or Enhancement). In the Experimental group, however, this number drops down to 36.8\%, with API-domain labels being the majority (42.7\%), followed by Code component labels (20.6\%). This difference in distributions alludes to the usefulness of the API-domain labels.  

\begin{figure}[!htb]
\centering
\includegraphics[width=.4\textwidth] {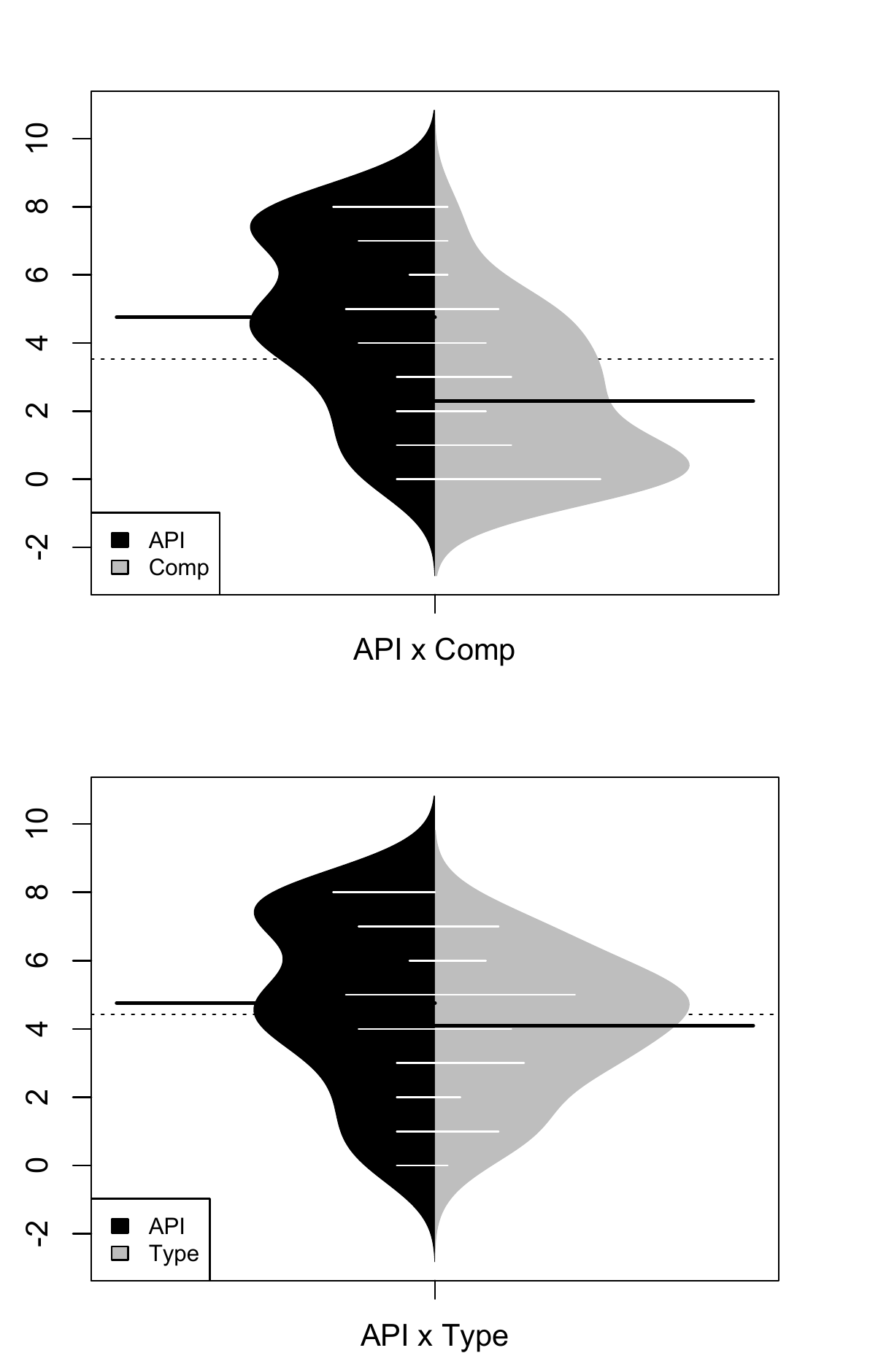}
\caption{Density Probability Labels (Y-Axis): API-domain x Components x Types.}
\label{fig:newlabels_countsAC}
\end{figure}

To better understand the usefulness of the API-domain labels as compared to the other types of labels, we further investigated the label choices among the Experimental group participants. Figure \ref{fig:newlabels_countsAC} presents two violin plots comparing (a) API-domain labels against Code component labels and (b) and type of issue. Wider sections of the violin plot represent a higher probability of observations taking a given value, the thinner sections correspond to a lower probability. The plots show that API-domain labels are more frequently chosen (median of 5 labels) as compared to Code component labels (median of 2 labels), with a large effect size ($|$d$|$ = 0.52). However, the distribution of the Issue Type and API-domain labels are similar as confirmed by a negligible effect size ($|$d$|$ = 0.1). These results indicate that the type of issue (bug fix, enhancement, suitable for newcomer) is important, as it allows developers to understand the type of task to which they would be contributing. Understanding the technical (API) requirements of solving the task is equally important in developers making their decision about which task to select.


Finally, we analyzed whether the demographic subgroups had different perceptions about the API-domain labels (Table~\ref{tab:apiXcompXtype}). When comparing Industry vs. Students, we found participants from industry selected 1.9x (p-value=0.001) more API-domain labels than students when we control by component labels. We found the same odds when we control by issue type (p-value=0.0007). When we compared Experienced vs. Novice coders, we did not find statistical significance (p=0.11) when controlling by components labels. However, we found that experienced coders selected 1.7x more API-domain labels than novice coders (p-value=0.01) when we control by type labels.

The odds ratio analysis suggests that API-domain labels are more likely to be perceived relevant got practitioners and experienced developers than by students and novice coders.

\begin{table}[h]
 \begin{center}
 \caption{Answers from different demographic subgroups regarding the API labels (API/Component/Issue Type)}
 \label{tab:apiXcompXtype}
 \begin{tabular}{c|c|r|r} 
  \hline
  \textbf{Subgroup} & \textbf{Comparison} & \textbf{API \%} & \textbf{Comp or Type \%} \\
  \hline

Industry & API/Comp & \textbf{56.0} & 44.0 \\
Students & API/Comp & 40.0 & \textbf{60.0} \\

Exp. Coders & API/Comp & \textbf{50.9} & 49.1 \\
Novice Coders & API/Comp & 41.5 & \textbf{58.5}  \\

Industry & API/issue Type & 45.5 & \textbf{55.5} \\
Students & API/issue Type & 30.6 & \textbf{69.4} \\

Exp. Coders & API/issue Type & 43.5 & \textbf{56.5} \\
Novice Coders & API/issue Type & 30.9 & \textbf{69.1}  \\

\hline
	
 \end{tabular}
 \end{center}
\end{table}

\MyBox{\textbf{\emph{RQ2 Summary.}} Our findings suggest that labels are relevant for selecting an issue to work on. API-domain labels increased the perception of the labels' relevancy. API-domain labels are specially relevant for industry and experienced coders.}

\section{Discussion}



\noindent\textbf{Are API-domain labels relevant?} 
Our results show that labels become more relevant for selecting issues when we introduce API-domain labels. From a new contributor's point of view, these labels are preferred over the current Code component labels. The fact that these labels are more generic than Code components, usually not exposing details about the technology, may explain the results. Future interviews can help investigate this topic.  

API-domain labels were shown to be more relevant for experienced coders. Additional research is necessary to provide effective ways to help novice contributors in onboarding. One could investigate, for example, labels that are less related to the technology and inform about difficult levels, priorities, estimated time to complete, contact for help, required/recommended academic courses, etc. 

Finding an appropriate issue involves multiple aspects, one of which is knowing the APIs required, which our labels are about. Our findings show that participants consider API-domain labels relevant in selecting issues. Future work can investigate if these labels lead to better decisions in terms of finding appropriate issues that match the developers' skills.

\noindent\textbf{Does the size of the corpus matter? } Observing the results reported for different corpora used as input, we noticed that the baseline model created using only the issue body had similar performance to the models using issue title, body, and comments or better performance than the model using only title. By inspecting the results, we noticed that by adding more words to create the model, the matrix of features becomes sparse and does not improve the classifier performance. 

\noindent\textbf{Domain imbalance and co-occurrence play a role.} We also found co-occurrence among labels. For instance, "UI", "Logging", and "IO" appeared together more often than the other labels. This is due to the strong relationship found in the source files. By searching the references for these API-domain categories in the source code, we found that "UI" was in 366 source code files, while "IO" and "Logging" was in 377 and 200, respectively. We also found that "UI" and "IO" co-occurred in 85 source files, while "UI" and "Logging" and "IO" and "Logging" co-occurred in 74 and 127 files, respectively. On the other hand, the API-domain labels for "Latex" and "Open Office Document" appeared only in five java files, while "Security" appears in only six files. Future research can investigate co-occurrence prediction techniques (e.g.,~\cite{wiese2017using}) in this context.

We suspect that the high occurrence of `UI'', ``Logging'', and ``IO'' labels ($>400$ issues) compared with the smallest occurrence of ``Security'', ``Open Office Documents'', ``Database'', ``PDF'', and ``Latex'' ($<32$ issues) may influence the precision and F-measure values. We tested the classifier with only the top 5 most prevalent API-domain labels and did not observe statistically significant differences. One possible explanation is that the transformation method used to create the classifier was Binary Relevance, which creates a single classifier for each label, overlooking possible co-occurrence of labels.

\noindent\textbf{The more specific the API-domain is, the harder it is to label.} Despite the lack of accuracy to predict the rare labels, we were able to predict those with more than 50 occurrences with reasonable precision. We argue that JabRef's nature contributes to the number of issues related to the "UI" and "IO." "Logging" occurs in all files and therefore explains its high occurrence. On the other hand, some specific API domains that are supposedly highly relevant to JabRef---such as ``Latex'', ``PDF'', and ``Open Office Documents''---are not well represented in the predictions. 

\noindent\textbf{How could we label issues with API domains that are rare in our dataset?} Looking to the Table~\ref{tab:confusionM} and comparing it with the aforementioned co-occurrence data, we can determine some expectations and induce some predictions. For example, the "database" label occurred with more frequency when we had "UI" and "IO". So, it is possible to guess when an issue has both labels, and we likely can suggest a "database "label, even when the machine learning algorithm could not predict it. The same can happen with the "Latex" label, which co-occurred with "IO" and "Network". A possible future work can combine the machine learning algorithm proposed in this work with frequent itemset mining techniques, such as apriori \cite{Apriori}. Thus, we can find an association rule between the previously classified labels.


\begin{table}[h!]
 \begin{center}
 \caption{overall metrics from the selected model}
 \label{tab:confusionM}
 \begin{tabular}{c|r|r|r|r|r|r} 
  \hline
  \textbf{API-Domain} & \textbf{TN} & \textbf{FP} & \textbf{FN} &
  \textbf{TP} & 
  \textbf{Precision} &
  \textbf{Recall} \\
  \hline

Google Commons & 107 & 15 & 27 & 30 & 66.6\% & 52.6\% \\
Test&112&18&29&20 & 52.6\% & 40.8\% \\
OS&152&8&8&11 & 57.8\% & 57.8\%\\
IO&9&30&3&137 & 82.0\% & 97.8\%\\
UI&30&26&10&113 & 81.2\% & 91.8\%\\
Network&107&10&30&32 & 76.1\% & 51.6\%\\
Security &167&6&2&4 & 40.0\% & 66.6\%\\
OpenOffice &165&6&3&5 & 45.4\% & 62.5\%\\
Database& 154&3&6&16 & 84.2\% & 72.7\%\\
PDF&164&5&4&6 & 54.5\% & 60.0\%\\
Logging&19&32&18&110 & 77.4\% & 85.9\%\\
Latex&170&1&1&7 & 87.5\% & 87.5\% \\
\hline
 \end{tabular}
 \end{center}
\end{table}

\noindent\textbf{What information is relevant when selecting an issue to contribute to?} Participants often selected \textsc{title}, \textsc{body}, and \textsc{labels} to look for information when deciding to which issue to contribute to. This result is in line with what we observed in the design of the machine learning algorithm.

\noindent\textbf{Practical implications} 
This research has implications for different stakeholders. 
    
\MyPara{New contributors.} API-domain labels can help open source contributors, enabling them to review the skills needed to work on the issues up front. This is specially useful for new contributors and casual contributors~\cite{Pinto:SANER:2016,AnitaCSCW}, who has no previous experience with the project terminology.

\MyPara{Project maintainers.} Automatic API-domain labeling can help maintainers distribute team effort to address project tasks based on required expertise. Project maintainers can also identify which type of APIs generate more issues in the project. Our results show that we can predict the most prominent API domains---in this case, "UI", "Logging", "IO", "Network", and "Test" with precision and recall up to 87.5\% and 97.8\%, respectively (see~Table \ref{tab:confusionM}).

\MyPara{Platform/Forge Managers.} Our results can be used to propose better layouts for the issue list and detail pages, prioritizing them against other information regions (\ref{fig:hotspotsurvey}). In the issue detail page on GitHub, for instance, the label information appears outside of the main contributor focus, on the right side of the screen. Moreover, as some wrong predictions in our study might be possibly caused by titles and body with little useful information to the corpus,  templates can guide GitHub users in filling out the issues' body to create patterns that help classifiers use the information to predict API labels. 


\MyPara{Researchers.} The scientific community can extend the proposed approach to other languages and projects, including more data and different algorithms. Our approach can also be used to improve tools that recommend tasks that match new contributor's skills and career goals (e.g., \cite{sarma2016training}).

\MyPara{Educators.} Educators who assign contributions to OSS as part of their coursework~\cite{pinto2017training} can also benefit from our approach. Labeling issues on OSS projects can help them select examples or tasks for their classes, bringing a practical perspective to the learning environment.

\section{Threats to Validity}

One of the threats to the validity of this study is the API domain categorization. We acknowledge the threat that different individuals can create different categorizations, which may introduce some bias in our results. To mitigate this problem, three individuals, including a Java Developer expert and a contributor to the JabRef project, created the API-domain labels. In the future, we can improve this classification process with (semi-)automated or collaborative approaches (e.g., \cite{8603296,lu2017learning}). 


Another concern is the number of issues in our dataset and the link between issues and pull requests. To include an issue in the dataset, we needed to link it to its solution submitted via pull request. By linking the issue with its correspondent pull request, we could identify the APIs used to create the labels and define our ground truth (check Section~\ref{sec:DataCollection}). To ensure that the link was correctly identified, we selected a random sample of 50 issues and manually checked for consistency. All of the issues in this validation set were correctly linked to their pull requests.

In prediction models, overfitting occurs when a prediction model has random error or noise instead of an underlying relationship. During the model training phase, the algorithm used information not included in the test set. To mitigate this problem, we also used a shuffle method to randomize the training and test samples. 

Although participants with different profiles participated in the experiment, the sample cannot represent the entire population and results can be biased. The experiment link randomly assigned a group to each participant. However, some participants did not finish the survey and the groups ended up not being balanced. Also, the way we created subgroups can introduce bias in the analysis. The practitioners' classification as industry and students were done based on the location of the recruitment and some students could also be industry practitioners and vice-versa. However, the results of this analysis were corroborated by the aggregation by experience level. 

Further, we acknowledge that we did not investigate if the labels helped the users to find the most appropriate tasks. It was not part of the user study to evaluate how effective the API labels were to find a match with user skills. This would only be possible if we had the users to work on the issues, which was not part of the experiment. Besides, we did not evaluate how False Positive labels would impact task selection or ranking. Our focus was on understanding the relevance that the API-domain labels have on the participants' decision. However, we believe the impact is minimal since in the three most selected issues, out of 11 recommendations only 1 label was a a false positive. Investigating the effectiveness API labels for skill matching and the problems that misclassification cause are potential avenues for future work.


Generalization is also a limitation of any case study. The outcomes could differ for other projects or programming languages ecosystems. We expect to extend this approach in that direction in future work. Nevertheless, the case study in a real world system showed how a multi-label classification approach can be useful for predicting API-domain labels and how relevant such a label can be to new contributors. Moreover, the API-domain labels that we identified can generalize to other projects that use the same APIs across multiple project domains (Desktop and Web applications). JabRef adopts a common architecture (MVC) and frameworks (JavaFX, JUnit, etc.), which makes it similar to a large number of other projects. As described by \citet{qiu2016understanding}, projects adopt common APIs, accounting up to 53\% of the APIs used. Moreover, our data can be used as a training set for automated API-domain label generation in other projects.


\section{Conclusion}

In this paper, we investigate to what extent we can predict API-domain labels. To do that, we mined data from 705 issues from the JabRef project and predicted 12 API-domain labels over this dataset. The model that was created using the Random Forest algorithm, unigrams, and data from the issue body offered the best results. The labels most present in the issues can be predicted with high precision.


To investigate whether API-domain labels are helpful to contributors, we built an experiment to present a mocked list of open issues with the API-domain labels mixed with the original labels. We found that industry practitioners and experienced coders selected API-domain labels more often than students and novice coders. Participants also preferred API-domain labels over Code component labels, which were already used in the project.  

This study is a step toward helping new contributors match their API skills with each task and better identify an appropriate task to start their onboarding process into an OSS project. For future work, we will explore new projects, a word embedding approach (Word2vec) with domain trained data to vectorise the issues, and investigate a unified API label schema capable of accurately mapping the skills needed to contribute to OSS projects.

\section*{Acknowledgment}

This work is partially supported by the National Science Foundation under Grant numbers 1815486, 1815503, 1900903, and 1901031, CNPq grant \#313067/2020-1. We also thank the developers who spent their time answering our survey.

\bibliographystyle{IEEEtranN}
\bibliography{msr}
\end{document}